\documentclass[twocolumn,showpacs,preprintnumbers,amsmath,amssymb,pre,nofootinbib]{revtex4-1}

\usepackage{color}    
\usepackage{graphicx}
\usepackage{dcolumn}
\usepackage{bm}
\usepackage{subfigure}
\usepackage{amssymb}
\usepackage{multirow}
\usepackage{amsmath}
\usepackage{braket}
\graphicspath{{plots/}}

\begin{document}

\title{\textit{Ab Initio} Path Integral Monte Carlo Results for the Dynamic Structure Factor of Correlated Electrons: From the Electron Liquid to Warm Dense Matter }

\author{T.~Dornheim$^{1}$, S.~Groth$^{1}$, J.~Vorberger$^{2}$, and M.~Bonitz$^{1}$}

\affiliation{
 $^1$Institut f\"ur Theoretische Physik und Astrophysik, Christian-Albrechts-Universit\"at zu Kiel,
 Leibnizstra{\ss}e 15, D-24098 Kiel, Germany\\
  $^2$Helmholtz-Zentrum Dresden-Rossendorf, D-01328 Dresden, Germany}

\begin{abstract}

The accurate description of electrons at extreme density and temperature is of paramount importance for, e.g., the understanding of astrophysical objects and inertial confinement fusion.
In this context, the dynamic structure factor $S(\mathbf{q},\omega)$ constitutes a key quantity as it is directly measured in X-ray Thomson (XRTS) scattering experiments and governs transport properties like the dynamic conductivity. In this work, we present the first \textit{ab initio} results for $S(\mathbf{q},\omega)$ by carrying out extensive path integral Monte Carlo simulations and developing a new method for the required analytic continuation, which is based on the stochastic sampling of the dynamic local field correction $G(\mathbf{q},\omega)$.
In addition, we find that the so-called static approximation
constitutes a promising opportunity to obtain high-quality data for $S(\mathbf{q},\omega)$ over substantial parts of the warm dense matter regime.

\end{abstract}

\maketitle

Over the recent years, there has been a remarkable spark of interest in so-called warm dense matter (WDM), an extreme state with high densities ($r_s=\overline{a}/a_\textnormal{B}\sim1$, $\overline{a}$ is the mean interparticle distance and $a_{\textnormal{B}}$ the Bohr radius) and temperatures ($\theta=k_\textnormal{B}T/E_\textnormal{F}\sim1$, with $E_\textnormal{F}=\hbar^2 q_\textnormal{F}^2/2m$ and $q_\textnormal{F}=(9\pi/4)^{1/3}a_\textnormal{B}/r_s$ being the Fermi energy and wave number). These conditions occur, for example, in astrophysical objects such as white and brown dwarfs~\cite{saumon1,saumon2,saumon3,becker,glenzer} and giant planet interiors~\cite{militzer1,militzer2,knudson,militzer3,manuel}, hot-electron chemistry~\cite{mukherjee,brongersma}, laser-excited solids~\cite{ernstorfer}, and along the compression path in inertial confinement fusion experiments~\cite{nora,schmit,hurricane3,kritcher}. WDM is nowadays routinely realized at large research facilities like NIF~\cite{nif1,nif2}, LCLS~\cite{lcls1,sperling}, and the European X-FEL~\cite{xfel1}.
Here X-ray Thomson scattering (XRTS)~\cite{siegfried_review,valenzuela,jan2} has emerged as an important method of diagnostics, with the electronic dynamic structure factor $S(\mathbf{q},\omega)$ being the central quantity. However, to make XRTS a reliable tool, an accurate theoretical description of the dynamic density response of warm dense 
electrons is indispensable~\cite{dominik}.

In this Letter, we focus on the uniform electron gas (UEG), one of the most fundamental model systems in physics and quantum chemistry~\cite{quantum_theory,loos}. While the static properties of the UEG in the ground state have 
mostly 
been known for over three decades~\cite{gs1,gs2,vwn,perdew}, the intricate interplay of 
Coulomb coupling and quantum degeneracy effects with thermal excitations
has rendered a thermodynamic description in the warm dense regime a challenging problem that has only been solved recently~\cite{schoof_prl,dornheim_prl,groth_prl}, see Ref.~\cite{review} for an extensive review. Naturally, \textit{dynamic} simulations of electrons that are required for frequency-resolved properties (dynamic conductivity, optical absorption, collective excitations etc.) and rigorously take into account all aforementioned effects are even more difficult. Therefore, results for $S(\mathbf{q},\omega)$ at WDM conditions that go beyond the random phase approximation (RPA)~\cite{pines} are sparse and have been obtained using 
uncontrolled 
approximations, such as  
diagram-summation-based Green function techniques~\cite{kwong,kas1,kas2,kas3}.
On the other hand, \textit{ab initio} path integral Monte Carlo (PIMC)~\cite{cep} simulations can provide an exact description, but are limited to static properties that can be formulated in terms of an ``imaginary time'', $i\tau \in [0,i \hbar \beta]$. Therefore, to obtain  quantities that depend on frequency, such as $S(\textbf{q},\omega)$,  one has to perform an analytic continuation from imaginary to real times. Unfortunately, this constitutes a notoriously difficult problem~\cite{jarrell,vitali,schoett}, and a universal approach is missing.

In this work, we overcome this difficulty for the case of the UEG. I) We carry out PIMC calculations of the imaginary-time density--density correlation function $F(\mathbf{q},\tau)$ [see Eq.~(\ref{eq:F})] for different temperatures $\theta=0.75, 1, 2, 4$, going from the WDM regime ($r_s=2$) to the strongly correlated electron liquid ($r_s=10$). II) These data serve as the main input for a new reconstruction method that 
 is based on a stochastic sampling of the dynamic local field correction (LFC) $G(\mathbf{q},\omega)$ [see Eq.~(\ref{eq:define_LFC})], which allows to fulfill a multitude of exact constraints. III) We are thereby able to present the first accurate results for the dynamic structure factor at WDM conditions and to explicitly study the combined impact of quantum diffraction, temperature and correlation effects on the dispersion relation. IV) We furthermore investigate the \textit{static approximation} and uncover that it yields highly accurate results in a broad parameter range. 
This opens new avenues towards \textit{ab initio} data for $S(\mathbf{q},\omega)$ at conditions that would otherwise be inaccessible
due to the fermion sign problem~\cite{dornheim_pop,troyer,loh}.

In addition to being interesting in their own right, we expect our accurate results for $S(\mathbf{q},\omega)$ of the UEG to be directly useful for, e.g., the interpretation of WDM experiments~\cite{dominik} (e.g., using the Chihara decomposition~\cite{chihara1,chihara2}), 
various density functional theory (DFT) applications~\cite{lu,patrick,burke2,gross,tddft}, and quantum hydrodynamics~\cite{zhandos}. Dynamic local field corrections are used not only for the calculation
of equilibrium properties but are an important input for several
non-equilibrium quantities like the stopping power
\cite{Cayzac:2017,Fu:2017}, for electron-ion energy transfer rates
\cite{Vorberger:2010}, for the electrical and thermal conductivities
\cite{Desjarlais:2017,Veysman:2016}, or for collisional absorption of
laser energy \cite{Hilse:2005}.

\begin{figure*}
\includegraphics[width=0.9\textwidth]{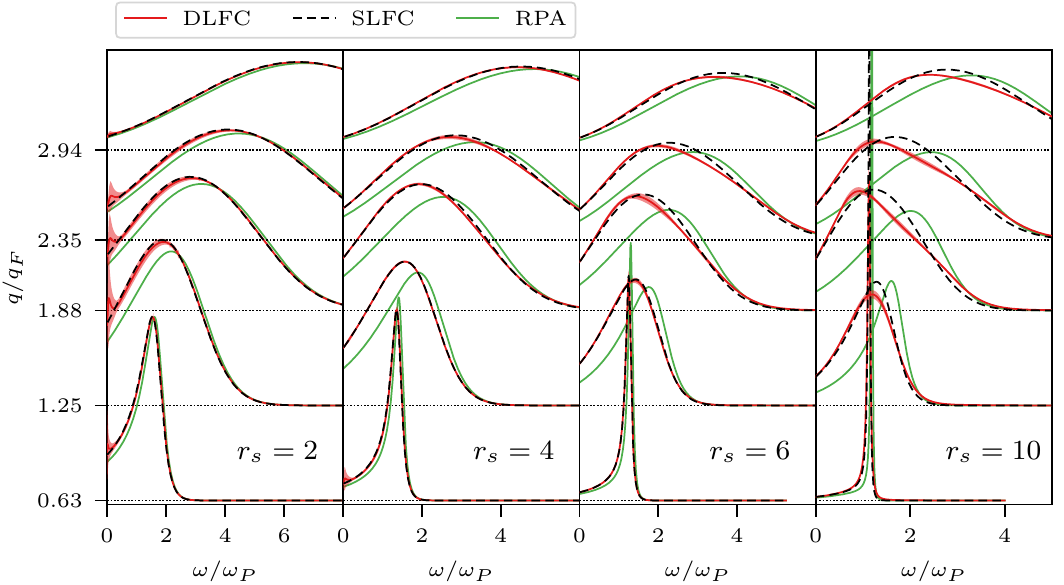}
\caption{\label{fig:panel}
Dynamic structure factor of the uniform electron gas at $\theta=1$ for different $r_s$-values. DLFC and SLFC correspond to the full reconstructed dynamic LFC (red) and to 
using
$G_{\rm static}(\mathbf{q},\omega)=\textnormal{Re}\,G(\mathbf{q},0)$ (black dashes), respectively. Note that the curves for different $r_s$ have been multiplied with scaling-factors. The shaded areas depict the given interval of uncertainty.
}
\end{figure*}  

\textbf{Theory:} We carry out fermionic PIMC simulations of the UEG in the canonical ensemble using a variation of the worm algorithm by Boninsegni \textit{et al.}~\cite{boninsegni1,boninsegni2} to compute the imaginary-time density--density correlation function (we use Hartree atomic units throughout this work)
\begin{eqnarray}\label{eq:F}
F(\mathbf{q},\tau) = \frac{1}{N} \braket{n_\mathbf{q}(\tau)n_{-\mathbf{q}}(0)}\ ,
\end{eqnarray}where the Fourier components of the density, $n_\mathbf{q}(\tau)$, are evaluated at 
imaginary times $\tau\in[0,\beta]$, see Refs.~\cite{berne1,berne2} for details and the Supplemental Material~\cite{supplement} for a graphical depiction. It should be noted that, although PIMC simulations of electrons are afflicted with a severe sign problem~\cite{troyer,dornheim_pop,loh,review}, which constitutes the main obstacle in our simulations, a straightforward evaluation of Eq.~(\ref{eq:F}) using the more advanced permutation blocking PIMC (PB-PIMC) and configuration PIMC (CPIMC) methods~\cite{dornheim,dornheim2,groth,dornheim3} is not yet possible.

Eq.~(\ref{eq:F}) is connected to the dynamic structure factor via 
\begin{eqnarray}\label{eq:F_S}
F(\mathbf{q},\tau) = \int_{-\infty}^\infty \textnormal{d}\omega\ S(\mathbf{q},\omega) e^{-\tau\omega}\ ,
\end{eqnarray}
which means that the task at hand is to perform an inverse Laplace transform to solve for $S(\mathbf{q},\omega)$.
In addition to our PIMC data for $F(\mathbf{q},\tau)$, it is also possible to obtain 
exact results for four frequency moments~\cite{quantum_theory,iwamoto,iwamoto2,kugler2}
\begin{eqnarray}\label{eq:k}
\braket{\omega^k} = \int_{-\infty}^\infty \textnormal{d}\omega\ \omega^k S(\mathbf{q},\omega)\ ,
\quad k=-1,0,1,3 \ .
\end{eqnarray}
see Ref.~\cite{supplement} for details.
The typical strategy would now be to find a trial function $S_\textnormal{trial}(\mathbf{q},\omega)$, which, when plugged into Eqs.~(\ref{eq:F_S}) and (\ref{eq:k}) reproduces the PIMC data within the given statistical uncertainty. We investigated different approaches, including the genetic evolution of an entire trial population~\cite{vitali} and the application of a deep neural network to \textit{learn} the needed inverse Laplace transform~\cite{deep_learning}. 
Unfortunately, the different methods did not converge 
towards the same solution for $S(\mathbf{q},\omega)$
in many cases, which means 
 that the information about the dynamic structure factor provided by Eqs.~(\ref{eq:F_S}) and (\ref{eq:k}) is not sufficient to determine a unique solution.

To overcome this obstacle, we make use of the fluctuation--dissipation theorem~\cite{kugler1,quantum_theory}
\begin{eqnarray}
S(\mathbf{q},\omega) = - \frac{ \textnormal{Im}\,\chi(\mathbf{q},\omega)  }{ \pi n (1-e^{-\beta\omega})}\ ,
\end{eqnarray}
which links $S(\mathbf{q},\omega)$ to the imaginary part of the density response function 
\begin{eqnarray}\label{eq:define_LFC}
\chi(\mathbf{q},\omega) = \frac{ \chi_0(\mathbf{q},\omega) }{ 1 - v_q\big[1-G(\mathbf{q},\omega)\big]\chi_0(\mathbf{q},\omega)}\ ,
\end{eqnarray}
with $v_q=4\pi/q^2$, and $\chi_0(\mathbf{q},\omega)$ 
refering to the noninteracting system. The dynamic LFC $G(\mathbf{q},\omega)\in\mathbb{C}$ contains all exchange-correlation effects beyond the mean field-level, i.e., the 
RPA is recovered by setting 
$G=0$.

Thus the computation of $S(\mathbf{q},\omega)$ has been reformulated into a quest for $G(\mathbf{q},\omega)$,
which is highly advantageous as many properties of the latter are known.
More specifically, we stochastically sample trial solutions $G_\textnormal{trial}(\mathbf{q},\omega)$ that exactly fulfill (i) the Kramers-Kronig relation between $\textnormal{Im}\,G(\mathbf{q},\omega)$ and $\textnormal{Re}\,G(\mathbf{q},\omega)$, (ii) that $\textnormal{Im}\,G(\mathbf{q},\omega)$  [$\textnormal{Re}\,G(\mathbf{q},\omega)$] is odd [even] with respect to $\omega$, 
(iii) the correct high and low frequency limits 
$\textnormal{Re}\,G(\mathbf{q},\infty)$ and $\textnormal{Re}\,G(\mathbf{q},0)$ known from our PIMC data, and (iv) the corresponding limits
$\textnormal{Im}\,G(\mathbf{q},\infty)=\textnormal{Im}\,G(\mathbf{q},0)=0$.
These $G_\textnormal{trial}(\mathbf{q},\omega)$ are then used to compute the corresponding trial structure factors, $S_\textnormal{trial}(\mathbf{q},\omega)$, which are subsequently plugged into Eqs.~(\ref{eq:F_S}) and (\ref{eq:k}) and discarded when they are not in agreement with the PIMC data for $F(\mathbf{q},\tau)$ and $\braket{\omega^k}$.

In contrast to the direct reconstruction of $S(\mathbf{q},\omega)$, the incorporation of the exact constraints on $G$ leads to a drastic reduction in the space of trial structure factors, and the problem becomes tractable. 
The final result is then computed as the average over the set of those $S_\textnormal{trial}(\mathbf{q},\omega)$ that reproduce $F(\mathbf{q},\tau)$ and $\braket{\omega^k}$ within the given statistical uncertainty. In addition, this allows us to compute the variance of this set as a measure of the remaining uncertainty, see 
Ref.~\cite{supplement} for details.

\begin{figure}
\includegraphics[width=0.45\textwidth]{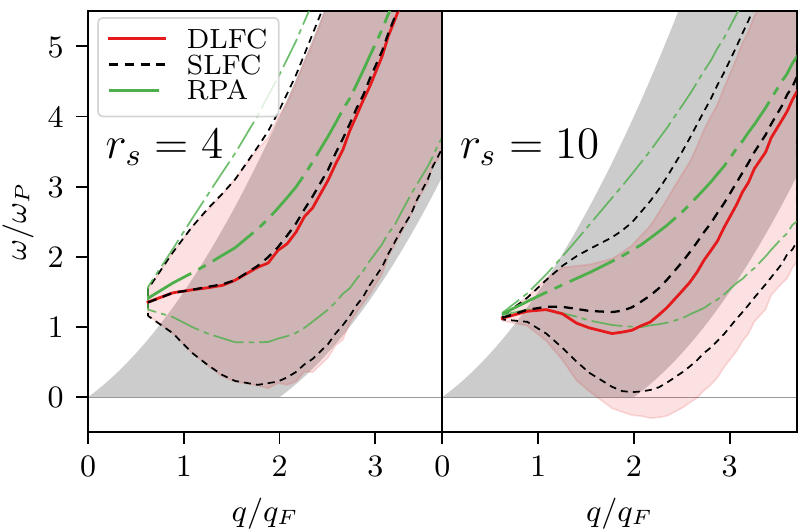}
\caption{\label{fig:FWHM}
Peak position (central lines) and full width at half maximum (red shaded area, for DLFC and outer lines, for SLFC and RPA) of $S(q,\omega)$ at $\theta=1$ for $r_s=4$ (left) and $r_s=10$ (right). The shaded grey area indicates the pair continuum in the ground state.
}
\end{figure}

\textbf{Results:} In Fig.~\ref{fig:panel}, we show our results for $S(\textbf{q},\omega)$
obtained for $N=34$ unpolarized electrons at $\theta=1$ for different values of $r_s$. To rule out possible finite-size effects, we have carried out PIMC simulations for larger particle numbers where possible. It turns out that, as in the case of the previously studied static structure factor~\cite{dornheim_prl,dornheim_pop,review,dornheim_cpp}, the only effect of the finite simulation box is the discrete set of available $\mathbf{q}$ values, but the functional form of $S(\mathbf{q},\omega)$ remains practically unchanged.
The green curves correspond to the RPA, which exhibits the well-known sharp plasmon peak below a critical wave vector $q_\textnormal{c}$ and gets significantly damped upon reaching the 
pair continuum~\cite{pines,quantum_theory}, cf.~Fig.~\ref{fig:FWHM} for the corresponding dispersion relation. 

The red curves depict the solutions of our stochastic dynamic LFC (DLFC) procedure. Observe the relatively large uncertainty in $S(\mathbf{q},\omega)$ for $r_s=2$ at low frequency, which is a direct consequence of the increased statistical uncertainty in $F(\mathbf{q},\tau)$ caused by the fermion sign problem. This allows for the possibility of a small diffusive peak~\cite{jan1}, which is most likely not a real physical effect, but cannot be ruled out on the basis of our PIMC data. Nevertheless, even at this high density, which falls well into the WDM regime, we are able to resolve significant deviations from the RPA around $q=2q_\textnormal{F}$.

With decreasing density, the deviations from the RPA curves become more pronounced, and we observe both a broadening and a significant red-shift at intermediate $q$ for $r_s=4$ and $r_s=6$. At the strongest coupling strength studied in this work, $r_s=10$, the DLFC curves 
exhibit a nontrivial superposition of 
a maximum at low $\omega$ and a broad shoulder that is following the peak in the RPA data. 
We stress that this behavior is caused by strong exchange-correlation effects in the dynamic density response and resembles ground state results by Takada \textit{et al.}~\cite{takada1,takada2} which were  interpreted as an incipient excitonic mode. 

Let us now consider the dashed black curves, which correspond to the \textit{static approximation} (SLFC), i.e., to setting $G_\textnormal{static}(\textbf{q},\omega) = \textnormal{Re}\,G(\mathbf{q},0)$ in Eq.~(\ref{eq:define_LFC}). This approach is motivated by the considerable success of static LFC-based schemes such as STLS (Singwi-Tosi-Land-Sj\"olander)~\cite{stls_original,stls,stls2} and VS (Vashishta-Singwi)~\cite{vs_original,stolzmann,stls2} in the description of the UEG, see Ref.~\cite{review} for a an extensive topical discussion. However, in contrast to the approximate treatment 
of $G_\textnormal{static}$ in those works, 
here we use the exact static limit of  $G$ that is computed from our PIMC data.
The results are striking: evidently, already the inclusion of the exact static LFC leads to an overall very good agreement with the exact results (except for $r_s=10$) and, thus, to a
remarkable 
improvement over the RPA. 
At $r_s=2$, 
no deviation from the DLFC curve can be resolved within the given accuracy. 
At $r_s=4$ and $r_s=6$, there appear small
deviations from the exact result
at intermediate $q$, but both the red shift and the broadening of the peak are reproduced. Even at $r_s=10$, 
the static approximation 
allows to 
captures the most important feature, 
i.e., the emergence of the low-frequency peak in $S(\mathbf{q},\omega)$.

In Fig.~\ref{fig:FWHM}, we show the corresponding dispersion relations derived from the peak of $S(\textbf{q},\omega)$ for $r_s=4$ and $r_s=10$. 
For the higher density, both the DLFC and SLFC curves exhibit a significant red shift and broadening compared to the RPA 
for $1\lesssim q/q_\textnormal{F}\lesssim 3$ whereas for small and large wave numbers all three curves eventually converge, as it is expected. Notice the striking agreement between DLFC and the static approximation both in peak position and width over the entire $q$-range.


At $r_s=10$ the exact dispersion relation exhibits an interesting non-monotonic behavior  with a minimum
around $q_\textnormal{min}\approx1.9q_\textnormal{F}$, which is in striking contrast to the monotonically increasing RPA curve. The minimum is also visible (although less pronounced) in the static approximation.
Such a \textit{negative dispersion} has previously been reported in the ground state both in experiments with alkali metals~\cite{vomFelde} and in static LFC-based calculations~\cite{taut1,fortm}, as well as in molecular dynamics simulations of the strongly coupled classical one-component plasma~\cite{ott}.

\begin{figure}
\includegraphics[width=0.45\textwidth]{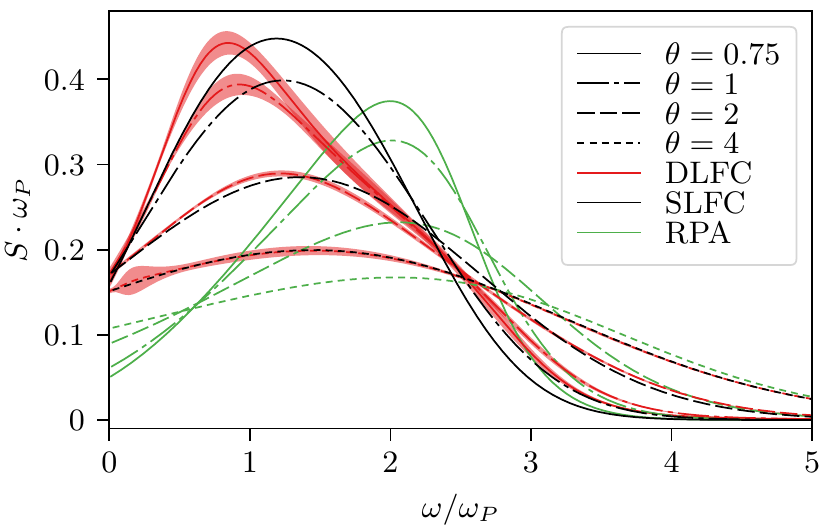}
\caption{\label{fig:T}
Dynamic structure factor 
for $r_s=10$,  $q/q_\textnormal{F}\approx1.88$, and different temperatures that are distinguished by the line styles. 
}
\end{figure}

Finally, in Fig.~\ref{fig:T}, we investigate the temperature dependence of $S(\mathbf{q},\omega)$ at $r_s=10$ for $q/q_\textnormal{F}\approx1.88$, i.e., in the minimum of the dispersion relation. 
At high temperature ($\theta=4$), DLFC and SLFC are practically indistinguishable whereas the RPA reproduces the broad peak qualitatively.
Decreasing the temperature to $\theta=2$ leads to a significantly increased deviation from RPA as the red shift becomes more pronounced. Still, this exchange-correlation effect is almost exactly captured by the static approximation. For the lowest temperatures, $\theta=1$ and $\theta=0.75$, the frequency dependence of $G(\mathbf{q},\omega)$ finally manifests itself, as the red curves exhibit a nontrivial shape with a low-frequency peak and a high-frequency shoulder, which get merged into a single broad peak in the static approximation.

\textbf{Summary and discussion:}
We have carried out PIMC simulations of the UEG from the electron liquid ($r_s=10$) to the WDM regime ($r_s=2$) for $0.75 \leq \Theta \leq 4$ and computed the imaginary-time density--density correlation function, $F(\mathbf{q},\tau)$, and various 
frequency moments, cf.~Eq.~(\ref{eq:k}). These data were used as input for our new reconstruction scheme to compute the first \textit{ab initio} data for the dynamic structure factor $S(\mathbf{q},\omega)$ at finite temperature without any approximation in the treatment of exchange-correlation, temperature, or quantum degeneracy effects. This was achieved by using the fluctuation-dissipation theorem to stochastically sample the dynamic local field correction, $G(\mathbf{q},\omega)$, fulfilling a multitude of exact properties, which renders the analytic continuation tractable.

This has allowed us to study the impact of correlation effects on $S(\mathbf{q},\omega)$, which 
manifest as a 
small, yet 
significant broadening and red shift 
at high density, and an emerging low frequency peak towards stronger coupling.
The latter is analogous to a possible incipient excitonic mode previously reported in the ground state~\cite{takada1,takada2,higuchi}.
The application of our approach to 
even lower densities, $10 <r_s<100$, to investigate a possible phase transition in the strongly correlated Fermi liquid~\cite{takada2} 
is beyond the scope of the present work and
remains a challenging topic for future research.

In addition to our full dynamic LFC, we have also studied the effect of the \textit{static approximation}, i.e., by setting $G_\textnormal{static}(\mathbf{q},\omega) = G(\mathbf{q},0)$
for all frequencies on $S(\mathbf{q},\omega)$. The results are very promising, as this approach constitutes a distinct improvement over the RPA for all considered parameters, in particular in the high density regime, $r_s \lesssim 4$, that is of interest in contemporary WDM research.
While the PIMC+DLFC calculations presented in this work are limited by the fermion sign problem, it was recently demonstrated~\cite{dornheim_pre,groth_jcp} that the simulation of the inhomogeneous electron gas using the novel CPIMC and PB-PIMC methods allows for accurate results of the static LFC over significant parts of the WDM regime. Therefore, a future extensive quantum Monte Carlo study of $G(\mathbf{q},0)$, which might culminate in a possible parametrization $G(\mathbf{q},0;r_s,\theta)$ analogous to previous ground state works~\cite{cdop,farid,utsumi}, 
is highly desirable.

We are confident that our \textit{ab initio} results for the dynamic density response of the UEG will be of high interest for the warm dense matter community and beyond. Direct applications include the interpretation of experiments using XRTS~\cite{siegfried_review} (see Ref.~\cite{dominik} for a topical discussion), the construction of novel functionals in DFT~\cite{lu,patrick,burke2} and time-dependent DFT~\cite{tddft}, and quantum hydrodynamics~\cite{zhandos}. In addition, our data will serve as a valuable benchmark for the development of new methods for the description of the dynamics of warm dense electrons, such as the method of moments by Tkachenko and co-workers~\cite{jan1,igor1,igor2,igor_quantum} and the recently presented advances in the nonequilibrium Green function method by Kas and Rehr~\cite{kas1,kas2}.



 \section*{Acknowledgments}
 T.D.~and S.G.~contributed equally to this work.
 We acknowledge fruitful discussions with H.~K\"ahlert and Zh.A.~Moldabekov.
This work has been supported by the Deutsche Forschungsgemeinschaft via project BO1366-10/2 and by the Norddeutscher Verbund f\"ur Hoch- und H\"ochleistungsrechnen (HLRN) via grant shp00015 for CPU time.

\end{document}